\documentclass[sigconf]{acmart}
\usepackage{graphicx} 
\usepackage[noend]{algpseudocode}
\usepackage{algorithm}
\usepackage{subcaption}
\usepackage{amsmath}
\usepackage{diagbox}
\usepackage{paralist}
\usepackage{amsmath}

\setcopyright{none} 
\renewcommand\footnotetextcopyrightpermission[1]{} 
\begin{document}

\title{Privacy-Preserving Identification of Target Patients from Outsourced Patient Data}


\author{Xiaojie Zhu}
\affiliation{%
  \institution{University of Oslo}
  \city{Oslo}
  \country{Norway}}
\email{xiaojiez@ifi.uio.no}

\author{Erman Ayday}
\affiliation{%
  \institution{Case Western Reserve University}
  \city{OH}
  \country{USA}}
\email{exa208@case.edu}

\author{Roman Vitenberg}
\affiliation{%
  \institution{University of Oslo}
  \city{Oslo}
  \country{Norway}}
\email{romanvi@ifi.uio.no}


\begin{abstract}

With the increasing affordability and availability of patient data,  hospitals tend to outsource their data to cloud service providers (CSPs) for the purpose of storage and analytics. However, the concern of data privacy significantly limits the data owners' choice. 
In this work, we propose the first solution, to the best of our knowledge, that allows a CSP to perform efficient 
identification of  target  patients (e.g., pre-processing for a genome-wide association study - GWAS)
over multi-tenant encrypted phenotype data (owned by multiple hospitals or data owners). 
We first propose an encryption mechanism for phenotype data, where each data owner is allowed to encrypt its data with a unique secret key. Moreover, the ciphertext supports privacy-preserving search and, consequently, enables the selection of the target group of patients (e.g., case and control groups). 
In addition, we provide a per-query based authorization mechanism for a client to access and operate on the data stored at the CSP. 
Based on the identified patients, the proposed scheme can either (i) directly conduct GWAS (i.e., computation of statistics about genomic variants) at the CSP or (ii) provide the identified groups to the client to directly query the corresponding data owners and conduct GWAS using existing distributed solutions. 
We implement the proposed scheme and run experiments over a real-life genomic dataset to show its effectiveness. 
The result shows that the proposed solution is capable to efficiently identify the case/control groups in a privacy-preserving way.

\end{abstract}


\maketitle 
 \pagestyle{plain}

\renewcommand{\algorithmicrequire}{\textbf{Input:}}
\renewcommand{\algorithmicensure}{\textbf{Output:}}

\section{Introduction}
\label{sec:paper3_introduction}



With the advent of precision medicine,
healthcare providers are empowered with the ability to select treatments based on a genetic understanding of the patient’s disease.
For instance, for cancer patients, a potential treatment may be a combination of surgery, chemotherapy, radiation, and immunotherapy, depending on the type of cancer and its stage.  Precision medicine can help decide on specific personalized treatment plans with certain drugs proving more effective treatment for specific genetic profiles.

However, to pave the way to such precision medicine, research over large volumes of patient data is required. This leads to the popularity of large-scale patient data sharing, such as Patient-Centered Clinical Research Network (PCORNet) \cite{selby2012patient} in the US, 
TranSMART \cite{Athey2013} in the EU, and the Global Alliance for Genomics and Health (GA4GH) \cite{GAGH}. These data-sharing systems are required to comply
with the increasingly stringent privacy regulations (e.g., HIPAA \cite{hipaa} or GDPR \cite{gdpr}).

Both centralized and decentralized approaches have been explored in this context.
In the centralized schemes \cite{kim2015private,sadat2017safety}, all data owners encrypt their data and outsource them to a common repository, where collective computation becomes possible. In the decentralized schemes \cite{froelicher2021truly,raisaro2018m}, all data owners store their data locally and run a distributed protocol to conduct the computation over all databases. 
Although both of these approaches provide the ability to conduct computation over combined data, both approaches assume that the client (who sends a query for the computation) already knows which database owners to query for the analysis. However, in real-life, there is a pre-computation step to identify which databases are utilized in the query processing. For instance, if a physician wants to identify similar patients to a given one based on the symptoms and then conduct some statistical tests on the identified patients, they should first contact all database owners to identify the useful databases. Similarly, to conduct a genome-wide association study (GWAS), a client first needs to identify which database owners have genomes belonging to the desired case and control group specifications.

In this work we aim to fill this gap. In general, we propose to develop efficient, privacy-preserving algorithms that can conduct this pre-computation (i.e., identification of target patients) at a centralized setting, at a CSP, over data from multiple database owners while also providing inherent access control. After the pre-computation step, we support either (i) computation of statistics over the identified records at the CSP (e.g., for simple statistical operations, such as calculation of minor allele frequencies and chi-square values) or (ii) utilization of other distributed solutions (e.g., \cite{froelicher2021truly,raisaro2018m}) between the data owners of the identified records (for more complex operations).

In the following, we discuss the main challenges of conducting large-scale privacy-preserving identification of target patients over multi-tenant patient data at the CSP.

\noindent\textbf{Encryption of data using unique cryptographic keys.} 
To protect data from a potentially curious CSP, each data owner (e.g., hospital) should encrypt its dataset before outsourcing. 
Moreover, each hospital should use its unique secret key for encryption. 
By doing so, hospitals can maintain data privacy even if some of the hospitals are corrupted. 
On the other hand, when each hospital encrypts its dataset with its own unique key, computation over a combined dataset becomes challenging.

\noindent\textbf{Efficient and privacy-preserving search over encrypted data across hospitals.} 
All the patient data should be encrypted against a potentially curious CSP.  
However,
the ciphertext should support secure search operation to facilitate selection of 
target patients (e.g., case and control groups for the association study) in a privacy-preserving and efficient way. 
Moreover, to avoid issuing a separate query for each hospital's dataset, searchability of the encrypted data should be supported across hospitals.

\noindent\textbf{Fine-grained access control of data.}
Since the data is stored at the CSP, the data owner loses  the direct control of the data. 
In order to guarantee that the data is accessed properly, the 
data owner needs to enforce an access policy over the outsourced data. 

In this work, we propose a privacy-preserving framework for
the identification of target patients over outsourced patient data. 
To the best of our knowledge, this is the first framework that tackles all the aforementioned challenges. 
In the proposed scheme, each hospital encrypts its own dataset with its unique key and outsources the storage and processing of the search operation over encrypted data. 
We mainly focus on identifying the target patients based on their encrypted phenotype data (which is typically the case in GWAS). 
For privacy of phenotype data, we encrypt them with a novel pairing-based encryption algorithm. 
The proposed encryption algorithm also provides the ability to efficiently search over encrypted phenotype data from several hospitals. 
This enables efficient identification of 
target group of patients (e.g., 
case and control groups) across hospitals. 
Furthermore, to let the phenotype data be properly accessible by legitimate/authorized clients, we introduce a fine-grained authorization scheme. For a single authorization request, each phenotype is considered as a unit of authorization.
We implement and evaluate the proposed scheme using a real-life genomic dataset. Our result shows that the proposed scheme requires less than 2 seconds to identify the case and control groups (each of size 50) over a dataset with 1052 patients.

The rest of the paper is organized as follows.
In the next section, we discuss the state of the art. 
In Section \ref{sec:paper3_background},
we introduce the primitives we used in this paper. In Section \ref{sec:paper3_models}, we present the models, including system model, data model, query model, and security model. 
Then, in  Section \ref{sec:paper3_proposed_scheme} we describe the proposed scheme in detail. In Section \ref{sec:paper3_privacy_analysis}, we provide the privacy analysis of the proposed solution. In Section \ref{sec:paper3_evaluation}, we evaluate the performance. 
Finally, we 
conclude the paper in Section~\ref{sec:conclusion}.

\section{Related Work}
\label{sec:paper3_related_work}

 
 With the increasing volume of patient data, there are two lines of research work to process the data. 
 The first approach is the centralized solution, which requires large amount of data to be stored in a single repository. 
 In order to match the requirement of storage and computation power, the CSP is the most popular central repository. 
 However, concerns about data privacy in cloud storage arise due to malicious attacks from inside and outside of the CSPs~\cite{clouddatabreach}. 
 To resolve the concern, Kim \textit{et al.} \cite{kim2015private} applied the BGV scheme \cite{gentry2012homomorphic} and YASHE scheme \cite{bos2013improved} to encrypt the patient data and conduct secure evaluation of $\chi^2$ distribution over the encrypted data. 
Sadat \textit{et al.} \cite{sadat2017safety} proposed a hybrid system called \emph{SAFETY}, which combines Paillier encryption scheme and Intel Software Guard Extensions (Intel SGX) to improve the efficiency of the chi-square test. 
 The second approach is the decentralized solution, where each participant stores its data locally. In order to inter-operate the patient data over multiple data providers,
  Kamm \textit{et al.} \cite{kamm2013new} proposed  to secretly share the sensitive data among several parties and compute GWAS over the distributed data. Similarly, Bogdandov \textit{et al.} \cite{bogdanov2014privacy} 
 adopted secret-sharing based techniques to implement a privacy-preserving framework for statistical analysis on federated datasets. 
Raisaro \textit{et al}  \cite{raisaro2018m} combines 
homomorphic encryption and obfuscation techniques 
to achieve privacy-preserving medical data sharing among many clinical sites. 
 Froelicher \textit{et al.} \cite{froelicher2021truly} proposed a multiparty homomorphic encryption algorithm. 
 Based on the algorithm, each party can store the data locally and be able to run analysis algorithms over all the participants' data without privacy violation. 
 
In the previous work \cite{zhu2019privacy}, we
proposed a scheme to find similar patients based on genomic makeup. In that scheme, with an assumption that all the data owners share a secret key, all the data owners build an index for their data using the secret key to support similar patient search. In this paper,  we attempt to remove not only the assumption, but also the index.

\noindent\textbf{Contribution.} 
Compared with the existing work, our main contributions are as follows: 
\begin{asparaenum}
    \item Our scheme supports multiple hospitals to outsource their data to the CSP and each of them encrypts its data with its own key while computation can be properly conducted over all the data. 
    \item Previous work assumes that the case and control groups for the association study are already known. 
    This cannot meet the requirement of dynamic selection of case/control groups. 
    In our paper, we propose a scheme that supports to dynamically select the case and control groups, which can be easily integrated to the existing solutions of association study. 
    \item  The authorization mechanism in the proposed scheme is designed based on per query request and it supports fine-grained authorization. 
\end{asparaenum}

\section{Background}
\label{sec:paper3_background}
 In this section, 
we first briefly introduce the background of genomics. 
Then,
we introduce symmetric bilinear groups applied in the proposed pairing-based encryption algorithm. 

\subsection{Genomics Background}
\label{sec:paper3_genomics}
The most common mutation in human population is called single nucleotide polymorphism (SNP). It is the variation in a single nucleotide at a particular position of the genome~\cite{risch2000searching}. There are about 5 million SNPs observed per individual and sensitive information about individuals (such as disease predispositions) are typically inferred by analyzing the SNPs. 
Two kinds of nucleotides (or alleles) are observed for each SNP: (i) major allele is the one that is observed with a high frequency and (ii) minor allele is the one that is observed with low frequency. The frequency of the minor allele in a given population is denoted as the minor allele frequency (MAF). Each SNP includes two nucleotides, one inherited from the father and the other one from mother. For simplicity, we represent the value of a SNP $i$ as the number of its minor alleles, and hence $SNP_i \in \{0, 1, 2\}$. A SNP is represented by an (ID, value) pair, where the ID is taken from a large standardized set of strings and the value is in $\{0, 1, 2\}$. In the following sections, if we mention a SNP (or SNPs) without mentioning the ID or value, we mean both parts.

\subsection{Symmetric Bilinear Groups}
\label{sec:symmtric_bilinear_groups}
Let $G$ be an additive group of prime order $p$ and $g_1 \in G$ be the generators of $G$ and $g_2 \in G$ is a random element from $G$. Let $e: G \times G \rightarrow G_T$ be a function which maps two elements from $G$ to an element in the target group $G_T$  having prime order $p$. The tuple $(G, G_T, p, e)$ is an symmetric bilinear group if following properties hold:  \\
(a) the group operations in $G$, $G_T$ are efficiently computable.  \\
(b) the mapping $e$ from $G$ to $G_T$ is efficiently computable.   \\
(c) the mapping $e$ is non-degenerate: $e(g_1, g_2) \neq 1$.\\
(d) the mapping $e$ is bilinear: for all $a,b \in \mathbb{Z}_p$, $e(g_1^a, g_2^b)=e(g_1,g_2)^{ab}$. 
Based on the symmetric bilinear group, we design an encryption algorithm to encrypt  phenotypes of patients.

\section{Models}
\label{sec:paper3_models}
In this section, we describe the system, data, query, and security models for our proposed scheme.  
We present the frequently used notation in  Table \ref{tab:notation}.

\begin{table}[h]
    \centering
    \caption{Frequently used notation}
    \begin{tabular}{|c|l|}
    \hline 
    $\lambda$ & the security parameter of the proposed scheme  \\
    \hline 
      $\tau$ & the number of hospitals \\ 
      \hline 
     $n$   & the number of phenotype attributes \\
     & for all the hospitals \\
     \hline 
     $m$ & the number of SNP identities for all the hospitals \\
     \hline 
     $t_i$ & the number of patients in hospital $i$ \\  
      \hline 
    $N$ & the size of case and control groups \\
    \hline 
    $\mathbf{A}$ & the global ordered set of phenotype attributes, $|A|=n$ \\
    \hline 
    \textit{P.name} & the attribute of a phenotype, \textit{P.name} $\in$ $\mathbf{A}$ \\
    \hline
    \textit{P.val} & the value of a phenotype,  \textit{P.val} $\in$ \{$0$, $1$\} \\
    \hline 
    
    \textit{SNP.ID} & the identity of a SNP \\
    \hline 
    \textit{SNP.val} & the value of a SNP \\ 
     \hline 
     $p_{i,j}$ & the pseudonym of patient $j$ in hospital $i$. \\
     \hline 
     $\mathbf{v}_{i,j}$ & the vector of SNP values for patient $j$ in hospital $i$
     \\
     \hline 
     $\Psi_i$ &  set of all patient records in hospital $i$ \\
     \hline 
    $\theta_{i,j}$  &  a record of phenotype of patient $j$ in hospital $i$,  \\
    & including the pseudonym and all phenotypes  \\
    & of patient $j$ \\

     \hline 
       $\vartheta_{i,j}$
      & a record of SNP data of patient $j$ in hospital $i$,  \\
      & 
      including pseudonym and set of pairs of  \\
      & \textit{SNP.id} and \textit{SNP.val}
      \\
     \hline 
     $\mathbf{C}_{i,\textit{SNP}}$ & the encrypted SNP dataset of hospital $i$ \\
     \hline 
     $\textit{Dict}_i^P$ & the encrypted phenotype dataset of hospital $i$ \\ 
     \hline 
    
    
      $K_i$ & the symmetric  key of hospital $i$,  \\
      & applied to encrypt/decrypt phenotype data \\
     \hline 
     $PK_i$ & the private key of hospital $i$, only revealed to a \\
     & legitimate client \\
     \hline 
     $SK_i$ & the master key of hospital $i$ \\
     \hline 
     $q_c$ & the query generated by client  \textit{c} \\
     \hline 
     $tk_{i,c,k}$ & the \textit{transform key} that hospital $i$ generates  for \\
     &  authorizing client \textit{c} to access the phenotype $k$ \\
     \hline 
     $H$ & a hash function \\
     \hline 
     $e$ & a bilinear mapping  \\
     \hline 
    \end{tabular}
    \label{tab:notation}
\end{table}

\subsection{System Model}
\begin{figure*}
    \centering
    \includegraphics[scale=0.7]{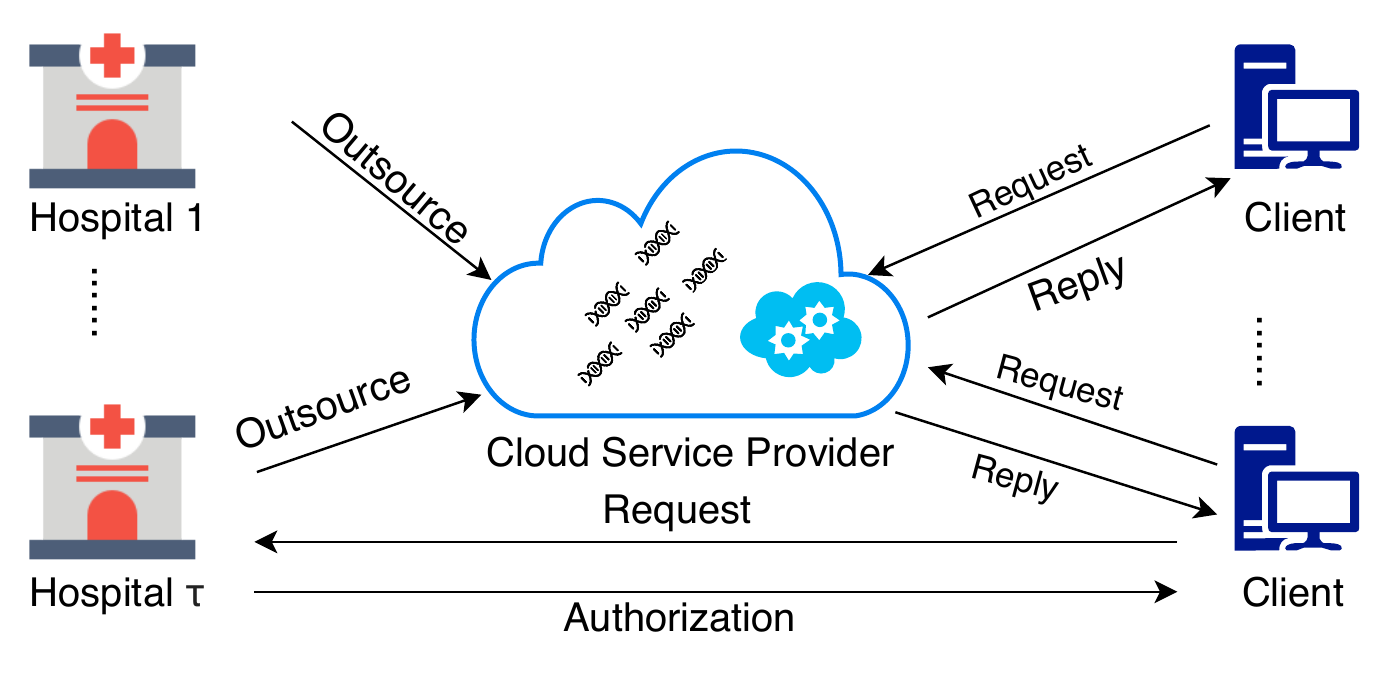}   
    \caption{Proposed system model. Hospitals encrypt their data and outsource them to the cloud service provider (CSP). A client sends an authorization request to a hospital for accessing its data. If the request is approved, the client gets authorization and is able to send request to the CSP to access the hospital's data.}
    \label{fig:paper3_system_model}
\end{figure*}

As shown in Figure~\ref{fig:paper3_system_model}, the system consists of three types of entities: clients, hospitals, and a cloud service provider (CSP). 
The hospital collects biological samples from patients and sequences them with patients' consent. 
In parallel, the hospital records various phenotypes of patients (e.g., height, eye color, and blood type). 
Instead of storing all genotype and phenotype data locally, the hospital encrypts the data and outsources them to the CSP. 
The client (e.g., a medical researcher) queries the CSP for different association studies on datasets of several hospitals. 
Before sending a query to the CSP, the client needs authorization from the involved hospital(s). 
If the client gets such an authorization, the corresponding computation is allowed to be conducted over all hospitals' datasets. 
Upon receiving a query from a client, the CSP first constructs the identification of target group of patients (e.g., case and control groups) by running the query over the encrypted phenotype data,
and then executes the other algorithms (e.g., GWAS) over the encrypted genotype data of individuals in the identified groups (e.g., case and control groups).

\subsection{Data Model}
In the proposed scheme, we make an assumptions to make sure that the data model is uniform across hospitals. 
We assume there exists a common set of terms applied to describe all the phenotypes across hospitals (referred as ``phenotype attributes''). That is, all the hospitals use the same terms to describe the same phenotypes.

For efficiency, we represent the value of a phenotype attribute in a binary format (as $0$ or $1$).  
$1$ means that the patient matches the phenotype attribute while $0$ denotes the opposite. 
For instance,
Table \ref{tab:paper3_phentype_data} illustrates a partial taxonomy of phenotypes that includes different height ranges in centimeters, as well as presence of breast cancer.

The value of each SNP is set to represent its number of minor alleles ($0$, $1$, or $2$) (see the details in Section \ref{sec:paper3_genomics}). Based on these settings, the phenotype and SNP dataset for hospital $i$ are represented as in Tables~\ref{tab:paper3_phentype_data} and~\ref{tab:paper3_snp_Data}, respectively. 
In the following sections, when we mention phenotype, it means both phenotype attribute and value, and when we mention SNP, it means both SNP identity and value.

\begin{table}[!htp]
    \centering
     \caption{ Phenotype dataset of hospital $i$ in terms of height and breast cancer.
      A patient record  includes a list of phenotype attribute values and each  of them is either $0$ or $1$.
     }
    \label{tab:paper3_phentype_data}
     \resizebox{\columnwidth}{!}{
    \begin{tabular}{|c|c|c|c|c|c|}
        \hline 
      \diagbox{Pseudonym}{P.name}  & \multicolumn{4}{c|}{height} & breast cancer \\
      \cline{2-5} 
      & $<100$ & [100,120) & $\cdots$
      & $>180$ &  \\
        \hline 
        $p_{i,1}$ & 0 &1 & $\cdots$ & 0 & 0 \\ 
        \hline 
        $\vdots$  & $\vdots$ & $\vdots$ & $\cdots$ & $\vdots$ &$\vdots$ \\
        \hline 
        $p_{i,t_i}$ & 0 & 0& $\cdots$ & 0 & 1\\ 
        \hline 
    \end{tabular}
     }
    
\end{table}


\begin{table}[!htp]
    \centering
    \caption{SNP dataset of hospital $i$.
     A patient record have a list of SNP values and each SNP value is either $0$, $1$ or $2$.
    }
    \label{tab:paper3_snp_Data}
    \begin{tabular}{|c|c|c|c|}
         \hline 
  \diagbox{Pseudonym}{SNP identity}
        & $\textit{SNP.ID}_1$ & $\cdots$ & $\textit{SNP.ID}_m$ \\
        \hline 
        $p_{i,1}$ & 1 & $\cdots$ & 0 \\ 
        \hline 
        $\vdots$ &  $\vdots$ & $\cdots$ &  $\vdots$ \\
        \hline 
        $p_{i,t_i}$ & 2 & $\cdots$ & 1 \\ 
        \hline 
    \end{tabular}
\end{table}

\subsection{Query Model}
The query mechanism is designed to find target groups of patients for specified phenotypes. 
In the proposed scheme, the query includes two parameters:
(i) a list of phenotypes of interest and  
(ii) a parameter to set the size of matching groups (e.g., case and control groups). 
As an instance, we use case and groups to illustrate the 
query mechanism. For simplicity, we assume the size of case and control groups to be equal, but it can be customized to support different sizes. 

Prior to outsourcing the data to the CSP, a hospital first encrypts its phenotype dataset by using the proposed pairing-based encryption algorithm that supports efficient search over encrypted phenotypes (as discussed in Section~\ref{sec:paper3_proposed_scheme}).
A query is generated by the client with input phenotype attributes and sent to the CSP. The proposed scheme allows a client to form the case and control groups based on multiple phenotype criterion (e.g., an association study for a particular type of cancer can be conducted only on males within a specific age range). 
The CSP, using the properties of the proposed pairing-based encryption, can check whether an encrypted patient's record (in a hospital's dataset) contains the phenotype attributes inside the query. If all the phenotype attributes in the query are included in a patient's record, the record is added into case group. In contrast, if a patient's record does not include any of the phenotype attributes in the query, the patient's record is added into the control group.  
Note that the query is encrypted without disclosing any phenotype information to the CSP and the CSP cannot learn any information from the search process. Thus, the CSP identifies the individuals in the case and control groups in a privacy-preserving way. 
The case and control groups may possibly contain patients from multiple hospitals. Specifically,
given a query from a client and \textit{transform key}s (detailed in Section~\ref{sec:paper3_clientauthoriatoin}) from hospitals that authorize the client to access their data, the CSP can search multiple hospitals' phenotype datasets (encrypted by using different secret keys). Thus, the search result is not limited to one hospital's dataset.

\subsection{Threat Model}
\label{sec:paper3_security_model}
Our threat model is consistent with previous work~\cite{lu2015privacy,schneider2019episode}.
Client, hospital(s), and the CSP are assumed to be semi-honest, that is, they honestly follow the protocol while trying to learn extra information during the protocol. 
A hospital may try to use its own data and knowledge to infer another hospital's data either via collusion with the CSP or exploring the common patient records in different hospitals. 
The CSP may analyze the stored ciphertext and observe the encrypted queries. Based on this information, the CSP may try to extract sensitive information. 
Also, a client may try to infer patients' data without having proper access authorization.  
Specifically, we consider following attacks: 

\noindent\textbf{Ciphertext analysis}: 
Since all the data are outsourced to the CSP, the CSP can
run different algorithms to analyze ciphertext and try to extract meaningful information. 

\noindent\textbf{Query analysis}: 
Since all the queries are sent to the CSP, the CSP may analyze the received queries and their frequency. 
Consequently, the CSP may try to infer the query pattern and content. 

\noindent\textbf{Operation analysis}:
The CSP conducts search computation and is able to obtain all the transcripts of the operation. Based on this information, the CSP may try to infer the content of query and stored ciphertext. 

\noindent\textbf{Unauthorized access}: 
Since all the hospitals' data is outsourced to the CSP, a client may try to access a hospital's data without authorization from that hospital. 

\noindent\textbf{Collusion between hospitals}: 
To infer a target hospital's data, several hospital may collude with each other and combine their knowledge. 

\noindent\textbf{Collusion between hospitals and the CSP}:
If some hospitals and the CSP reach consensus on common interest to learn another hospital's (victim's) data, all the related parties combine their knowledge and try to infer the sought information. 
For instance, if the search (at the CSP) includes two hospitals and if one of these hospital collude with the CSP, the CSP can learn which patients of the other (victim) hospital has the considered phenotype as a result of the search operation. However, to provide the common search functionality, this attack is unavoidable, and hence we do not consider it in this work. 

We thoroughly analyze all these attacks and robustness of the proposed scheme against them in Section~\ref{sec:paper3_privacy_analysis}. 
 
\section{Proposed Scheme}
\label{sec:paper3_proposed_scheme}
In this section, we first give an overview of the proposed scheme, and then describe its details.


\subsection{Overview}
In the proposed scheme, hospitals independently encrypt and outsource their phenotype  datasets to a CSP and the CSP conduct search operation over the outsourced federated data (from multiple hospitals) to identify target group of patients. 
To process phenotype data in an efficient and privacy-preserving way, we propose a novel encryption mechanism to encrypt the phenotype data. 
The proposed encryption scheme allows different hospitals to encrypt their phenotype data independently with their own secret keys. Moreover, the encryption algorithm supports identification of case and control groups efficiently, without information leakage. 
Furthermore, the identification process requires less communication compared with secure multi-party computation-based approaches~\cite{schneider2019episode} and less computation compared to homomorphic encryption-based approaches~\cite{akavia2019setup}. 
In general, the execution of privacy-preserving identification of target group of patients can be divided into seven phases:
data preprocessing, initialization, key generation, data encryption, client authorization, query generation, identification of case and control groups.  
We now present a high-level overview of these seven phases while the rest of this section provides in-depth details.

\noindent\textbf{Data Preprocessing.} Phenotype and genotype data need to be properly encoded in the required data format so that further processing can be conducted. 

\noindent\textbf{Initialization.} In this phase, the \emph{Setup} 
is performed to initialize the parameters and functions. 

\noindent\textbf{Key generation.}  In this phase, the \emph{KeyGen} function is  
executed to generate secret keys for hospitals. 

\noindent\textbf{Data encryption.}  
Hospitals recursively call \emph{SinglePhenotypeEncrypt} to encrypt their phenotype data. 

\noindent\textbf{Client authorization.} Before requesting access to a hospital's dataset (from the CSP), a client obtains an authorization from the hospital. 

\noindent\textbf{Query generation.} To run query over outsourced data with specified phenotypes, a query is generated by recursively running query generation algorithm \textit{SinglePhenotypeQueryGen} with the input of target phenotypes. 

\noindent\textbf{Identification of the case and control groups.} Given a query, the CSP identifies whether a patient record is in case group or control group by running the \textit{Search} algorithm.  




\subsection{Data Preprocessing}
\label{sec:paper3_preprocessdata}
In this section, we present the procedures we use to preprocess the 
phenotype data, so that the processed data matches the data format requirements. 

\subsubsection{Preprocessing Phenotype Data}
The representations of phenotype attributes should be uniform for all the hospitals, and hence we assume that all the hospitals share a common ordered set of phenotype attributes, denoted as $\mathbf{A}$. 
We also assume that the value corresponding to a phenotype attribute is binary: $1$ represents a patient having such phenotype attribute, while $0$ means the opposite.
In Table \ref{tab:paper3_phentype_data}, we illustrate the processed phenotype data for height and breast cancer. 
As seen in the table, in the dataset ($\Psi_i$) of hospital $i$, a patient record can be represented as $\theta_{i,j}=\{ p_{i,j}, (\textit{P.name}_1, \textit{P.val}_{j,1}), \cdots, (\textit{P.name}_n, \textit{P.val}_{j,n})\}$, where 
$\textit{P.name}_k \in \mathbf{A}$, 
$\textit{P.val}_{j,k} \in \{0,1\}$, $k \in [1,n]$ and $j \in [1, t_i]$. 


\subsection{System Initialization}\label{sec:paper3_initialization}
In this section, we present the procedures during the initialization so that all algorithms use the same initial parameters.

\subsubsection{Initialization for Phenotype Data Encryption}
With the input of the security parameter $\lambda$, the system first generates a symmetric bilinear mapping $e: G\times G \rightarrow G_T$, where the multiplicative cyclic group $G$ is generated by generator $g$ and has the prime order $p$ ($p>2^{\lambda}$). 
Then, a cryptographic hash function $H$ is selected. Above procedures are  detailed in 
\emph{Setup} algorithm that is shown in Algorithm \ref{alg:paper3_initialization}. 

\alglanguage{pseudocode}
 \begin{algorithm}
  \caption{Setup}
 \label{alg:paper3_initialization}
  \small
\begin{algorithmic}[1]
\Require $\lambda$
\Ensure $e$, $H$
\State set $e: G \times G \rightarrow G_T$ 
\State choose $H: \{0,1\}^*\times \{0,1\}^* \rightarrow
\{0,1\}^\lambda$ 
\State \Return $e$, $H$
\end{algorithmic}
\end{algorithm}



\subsection{Key Generation}
\label{sec:paper3_keygen}
In this section, we present the procedures to generate the required keys in the system.

\subsubsection{Key Generation for Phenotype Data Encryption}
Hospital $i$ randomly selects a master key $SK_i \in  Z_p$, and sets its private key as $PK_i=g^{SK_i} \in G$. 
In addition, hospital $i$ selects a secret key $K_i$  from $\{0,1\}^{\lambda}$.  
This is detailed in the \emph{Keygen} procedure presented in Algorithm \ref{alg:paper3_keygen_pairing}. 

 \alglanguage{pseudocode}
 \begin{algorithm}
  \caption{KeyGen}
 \label{alg:paper3_keygen_pairing}
  \small
\begin{algorithmic}[1]
\Require $i$, $\lambda$ 
\Ensure  $PK_i$, $SK_i$, $K_i$
\State $SK_i \leftarrow Z_p$ 
\State $PK_i \leftarrow g^{SK_i}$ 
\State $K_i \leftarrow \{0,1\}^{\lambda}$ 
 \State \Return $PK_i$, $SK_i$, $K_i$
\end{algorithmic}
\end{algorithm}


\subsubsection{Key Generation for client}
The system  randomly selects  a number $SK_c \in \mathbb{Z}_p$ and generates a key $PK_c=g^{SK_c}$ for each client. Once a client joins the system, it is assigned $SK_c$ and $PK_c$.

\subsection{Data Encryption  }\label{sec:paper3_dataencryption}
In this section, we  describe the proposed phenotype data encryption algorithm, which supports efficient privacy-preserving search and update. 

\subsubsection{Phenotype Data Encryption}
\label{sec:paper3_dataencryption_phenotypeenc}
The set of phenotype attributes is denoted by $\mathbf{A}$, as described before.
Phenotype data of a patient $j$ in hospital $i$ is represented as $\theta_{i,j} = \{p_{i,j}$, 
($\textit{P.name}_1$, $\textit{P.val}_{j,1}$), $\cdots$, ($\textit{P.name}_n$, $\textit{P.val}_{j,n})\}$. 
For each phenotype $k$ of the patient (phenotype name-value pair, i.e., $\textit{P.name}_k, \textit{P.val}_{j,k}$), the hospital first selects a random number $r_{i,j,k}$ from $\mathbb{Z}_p$, where $p$ is a large prime that is larger than $2^{\lambda}$. The pair of phenotype attribute and value is first hashed and then encrypted into a group value $c_{i,j,k}$. Afterwards, a symmetric encryption algorithm \textit{SE} (e.g., AES) is invoked with the input of a secret key $K_i$ and the pair of phenotype attribute $\textit{P.name}_k$ and value $\textit{P.val}_{j,k}$. 
Finally, the algorithm outputs the ciphertext $\hat{c}_{i,j,k}$ consisting of a random group element $g
^{r_{i,j,k}}$, an encoded group element  $c_{i,j,k}$, and a symmetrically encrypted ciphertext $\bar{c}_{i,j,k}$.

Algorithm \ref{alg:paper3_encrypt_pairing} shows how hospital $i$ encrypts a single phenotype $k$ of patient $j$ (i.e., $\textit{P.name}_k, \textit{P.val}_{j,k}$). 
To encrypt all patients' phenotype data in a hospital, we iteratively encrypt 
each patient's phenotypes. In detail, 
for each patient $j$ in hospital $i$, the
hospital first reads its phenotype data $\theta_{i,j}$, and then invokes the \textit{SinglePhenotypeEncrypt} to encrypt each phenotype of the patient. Once all the phenotypes of the patient $j$ is encrypted, the result denoted as $\mathbf{C}_{i,j}$ is stored in the dataset 
 $\textit{Dict}_{i}^P$ with the patient pseudonym $p_{i,j}$.  This process is repeated for each patient, detailed in Algorithm \ref{alg:paper3_edbencrypt_pairing}.

 \alglanguage{pseudocode}
 \begin{algorithm}
  \caption{SinglePhenotypeEncrypt}
 \label{alg:paper3_encrypt_pairing}
  \small
\begin{algorithmic}[1]
\Require ($\textit{P.name}_{k}$, $\textit{P.val}_{j,k}$)$\in \theta_{i,j}$,  $SK_i$, $K_i$
\Ensure $\hat{c}_{i,j,k}$ 
\State    $r_{i,j,k} \xleftarrow{\$} \mathbb{Z}_p$ 
\State    $\alpha_{i,j,k} \leftarrow H(\textit{P.name}_{k},\textit{P.val}_{j,k})$ 
\State    $c_{i,j,k} \leftarrow g^{-r_{i,j,k}/(SK_i-\alpha_{i,j,k})}$ 
\State  $\bar{c}_{i,j,k} \leftarrow SE(K_i,\textit{P.name}_{k} \circ \textit{P.val}_{j,k} )$ 
\State $\hat{c}_{i,j,k} \leftarrow (g^{r_{i,j,k}}, c_{i,j,k}, \bar{c}_{i,j,k})$ 
\State \Return $\hat{c}_{i,j,k}$
\end{algorithmic}
\end{algorithm}

\alglanguage{pseudocode}
 \begin{algorithm}
  \caption{PhenotypeEnc}
 \label{alg:paper3_edbencrypt_pairing}
\begin{algorithmic}[1]
\Require $\Psi_i$, $SK_i$, $K_i$
\Ensure $\textit{Dict}_i^P$
\State initialize a dictionary $\textit{Dict}_i^P$ 
\ForAll{$\theta_{i,j} \in \Psi_i$}
       
        \State $\mathbf{C}_{i,j} \leftarrow \phi$
        \ForAll{ $(\textit{P.name}_k, \textit{P.val}_{j,k}) \in \theta_{i,j}$}
        \State $\hat{c}_{i,j,k}\leftarrow \textit{SinglePhenotypeEncrypt}(\textit{P.name}_k, \textit{P.val}_{j,k}, SK_i, K_i)$
        \State $\mathbf{C}_{i,j} \leftarrow \mathbf{C}_{i,j} \cup \hat{c}_{i,j,k}$
        \EndFor
        \State $p_{i,j} \leftarrow \theta_{i,j}$
        \State $\textit{Dict}_{i}^P[p_{i,j}]  \leftarrow \mathbf{C}_{i,j}$
\EndFor
\State \Return  $\textit{Dict}_{i}^P$
\end{algorithmic}
\end{algorithm}

\subsection{Client Authorization}\label{sec:paper3_clientauthoriatoin}
In the proposed scheme, a client needs to get authorization from a hospital before it can access to (operate on) a hospital's data. 
Meanwhile, the hospital is capable to authorise the client in fine-granularity and the CSP should not learn the data that the hospital authorises the client to access. 
 In this section, we first present a simple client authorization mechanism, which requires a client to generate a query for each  hospital that authorizes the client to access its data. 
 In addition, this mechanism allows a client to access a hospital's data without any limitations once it is authorized.
 Then, to achieve flexible authorization and let the client generate a single query that can be used to operate on all authorized hospitals' datasets, we  further present an improved mechanism. The improved mechanism supports per-query based fine-grained authorization and it allows a single query to access multiple hospitals' data.

 \subsubsection{Simple Authorization}

In the simple authorization mechanism, if a client gets authorization from a hospital, the client can access all the data of the corresponding hospital for all future queries. 
To present this authorization mechanism, we assume that there exists a client with private key $PK_c = g^{SK_c}$ who wants to access (operate on) data from hospital $i$. 
The client first sends an authorization request to hospital $i$ with its 
private key $PK_c$. If the hospital approves the request, the hospital signs the private key $\sigma_{i,c}=PK_c^{SK_i}$ and sends it back to the client. The client recovers $PK_i = \sigma_{i,c}^{1/{SK_c}}$.
Upon obtaining $PK_i$, the client can generate a legitimate query. 
The details of this authorization mechanism are also shown in Algorithm~\ref{alg:paper3_authorisation_pairing}.

\alglanguage{pseudocode}
 \begin{algorithm}
  \caption{SimpleAuthorization}
 \label{alg:paper3_authorisation_pairing}
  \small
\begin{algorithmic}[1]
\Require $PK_c$, $SK_i$, $SK_c$
\Ensure $PK_i$
\State Client: send $PK_c = g^{SK_c}$ to hospital $i$ 
\State Hospital $i$: compute $\sigma_{i,c} \leftarrow PK_c^{SK_i}$ and send $\sigma_{i,c}$ to the client 
\State Client: compute $PK_i \leftarrow \sigma_i^{1/{SK_c}}$
\end{algorithmic}
\end{algorithm} 

\subsubsection{Improved Authorization}
\label{sec:paper3_clientauthoriatoin_impauthorization}
 
In the simple authorization mechanism, a client can access hospital's data indefinitely once the hospital authorizes the client. Also, the hospital is unable to control the granularity of the authorization. In other words, the hospital authorizes either all its data to a client or  none of them. 
Furthermore, simple authorization mechanism results in a query that cannot be executed across multiple hospitals' data (i.e., the client needs to generate separate queries for each hospital's dataset). 

In order to overcome these concerns, we propose an improved authorization mechanism. 
The new authorization mechanism supports fine-grained authorization on a per-query basis. Moreover, using the \emph{transform key} function, a single query sent to the CSP can be executed across hospitals. 
 As a result, the complexity of a client's query generation is reduced from O($\tau$) to $O(1)$, where $\tau$ is the number of accessible hospitals. 
The steps for the improved authorization mechanism are as follows. A client first generates an authorization request and sends it to a hospital.  
If the hospital approves the request, the hospital first generates a \emph{transform key} for the client (per query) and sends it to the CSP, such that the CSP can apply the \emph{transform key} to transform the ciphertext into the format that supports the client's query. After the transformation, the query can be executed by the CSP over the phenotype data of each hospital that authorizes the client. 
The construction of the \textit{transform key} needs to consider the privacy of the \textit{transform key}, transformation process, and the query. With these privacy requirements in mind, we set the \emph{transform key} at the granularity of phenotypes 
and compute it as  $tk_{i,c,k}=g^{\frac{-r_{c,k}(SK_i-SK_c)}{SK_i-\alpha_{c,k}}}$, where $i$ denotes the target hospital, $c$ represents the associated data from a client, and
$k$ represents the phenotype attribute in phenotype attribute set $\mathbf{A}$. 
The details of the \textit{transform key} construction are as follows.
\begin{enumerate}
    \item A client randomly selects $r_{c,k} \in \mathbb{Z}_p$~  ($1 \leq k \leq n$)  and sends it to a  hospital  with $PK_c$ and $\alpha_{c,k}$ ($1 \leq k \leq n$) to request corresponding data access.

    \item Upon receiving the query request, the hospital decides which phenotype attributes can be accessed and generates $tk_{i,c,k}$ ($1 \leq k \leq n$) for each approved phenotype attribute. Specifically, if the hospital approves the requested $\alpha_{c,k}$, the corresponding $tk_{i,c,k}$ is properly constructed based on $\alpha_{c,k}$ and $r_{c,k}$. Otherwise, a random group element is selected, such that the access structure can be protected from the CSP. 
    
    \item For each phenotype attribute,
    a transform key  $tk_{i,c,k}$ ($1 \leq k \leq n$)) is generated. The hospital sends all of them to the CSP and replies to the client with a \emph{success} message. 
\end{enumerate}

Given the transform key, the CSP can transform the ciphertext into the format that supports corresponding client's query. Using this technique, the CSP can process all the ciphertext before a query is launched. The details of the transformation are discussed in Section~\ref{sec:paper3_identificationcaseandcontrolgroup}.

\subsection{Query Generation}
\label{sec:paper3_querygen}
The query is generated based on client's input that specifies phenotype data of interest. Then, the query is executed over encrypted phenotype data. 
Algorithm~\ref{alg:paper3_querygen_pairing} shows query generation based on the input of a single phenotype.  
The client first applies the hash function $H$ on the input phenotype attribute and value, obtaining $\alpha_{c,k}$ ($k \in \mathbf{A}$) as the output.
Then, the client computes a group element $g^{r_{c,k}}$, where $r_{c,k}$ is selected during the generation of authorization request. $g^{r_{c,k}}$ is included in the query. Afterwards, the client computes the other part of the query:
$g^{-r_{c,k} \alpha_{c,k}}PK_c^{r_{c,k}}$. 
To support multiple phenotypes in a query, the client can iteratively invoke the SinglePhenotypeQueryGen algorithm. In detail, given a set $\Delta$ of phenotypes, for each phenotype inside $\Delta$, the client calls SinglePhenotypeQueryGen to encrypt it. The result $q_{c,k}$ ($1 \leq k \leq n$) is stored in a vector $\mathbf{v}_{c}$, which has the same dimension and the same order of phenotype attributes as $\mathbf{A}$. For the phenotypes whose phenotype attributes are in $\mathbf{A}$ but not in $\Delta$, the corresponding value is set to $0$ in vector $\mathbf{v}_c$. 
$q_c$ is a set and it includes the vector $\mathbf{v}_{c}$. 
Finally, in addition to $\mathbf{v}_c$, the size $N$ of case and control groups is also included in $q_c$ before $q_c$ is sent to the CSP. 

Algorithm~\ref{alg:paper3_squerygen_pairing} shows how to extends the input from a single phenotype to multiple phenotypes. 
Here, for each phenotype inside the input phenotype data set $\Delta$, the client invokes Algorithm~\ref{alg:paper3_querygen_pairing}. Once all the input data are processed, the algorithm outputs the final query $q_c$. 
Note that all the positions, where phenotype attributes are not included in the input phenotype data are filled with zeros. 
For example, a client can input a set of pairs of phenotype attribute and value as 
\{(breast cancer, 1), (lung cancer, 0), 
(blue eye color, 1)\} to generate a query.
Applying above algorithms, each target phenotype attribute is corresponding to a component $q_{c,i}$ in the query, the remaining that is not included in the input phenotype attribute set is set to $0$.
The final query also includes the size $N$ of case and control groups.

\alglanguage{pseudocode}
 \begin{algorithm}
  \caption{SinglePhenotypeQueryGen}
 \label{alg:paper3_querygen_pairing}
  \small
\begin{algorithmic}[1]
\Require{$PK_c$,  ($\textit{P.name}_k, \textit{P.val}_{c,k}$), $r_{c,k}$}
\Ensure{$q_{c,i}$}
\State        $\alpha_{c,k} \leftarrow H(\textit{P.name}_k, \textit{P.val}_{c,k})$ 
\State        $q_{c,k} \leftarrow (g^{r_{c,k}}, g^{-r_{c,k}\alpha_{c,k}}PK_c^{r_{c,k}})$
\State \Return $q_{c,k}$ 
\end{algorithmic}
\end{algorithm}


\alglanguage{pseudocode}
 \begin{algorithm}
  \caption{QueryGen}
 \label{alg:paper3_squerygen_pairing}
\begin{algorithmic}[1]
\Require $PK_c$, $\Delta$
\Ensure $q_{c}$
\State $q_c \leftarrow \phi$ 
\ForAll{$(\textit{P.name}_k,\textit{P.val}_{c,k}) \in  \Delta$}
\State  $q_{c,k} \leftarrow \textit{SinglePhenotypeQueryGen}(PK_c,\textit{P.name}_k,\textit{P.val}_{c,k})$ 
\State $q_c \leftarrow q_c \cup q_{c,k}$ 
\EndFor
\State fill $q_c$ with $0$ in locations of phenotype attributes in $A \backslash \Delta$ 
\State $q_c \leftarrow N$ 
\State \Return $q_c$
\end{algorithmic}
\end{algorithm}



\subsection{Identification of Case and Control groups}
\label{sec:paper3_identificationcaseandcontrolgroup}
Without loss of generality, we present the identification process at the CSP over a single hospital $i$'s dataset. Multiple hospitals' datasets are processed separately and in parallel.  
We describe the identification process in two steps. First, we show the search operation over a single phenotype.
Then, we extend the algorithm to multiple phenotypes. 
The details of search over a single phenotype are shown in Algorithm \ref{alg:paper3_search_pairing}. Specifically, 
given the ciphertext $\hat{c}_{i,j,k}$  of phenotype $k$ of patient $j$ in hospital $i$, 
we first extract the components  
$ g^{r_{i,j,k}}$, $c_{i,j,k}$, $\bar{c}_{i,j,k}$ from $\hat{c}_{i,j,k}$, denoted as $\hat{c}_{i,j,k,0}$, $\hat{c}_{i,j,k,1}$, and $\hat{c}_{i,j,k,2}$.
Based on the input of the query $q_{c,k}$ from client $c$ querying for phenotype $k$, we extract 
$ g^{r_{c,k}}, g^{-r_{c,k}\alpha_{c,k}}PK_c^{r_{c,k}}$
from $q_{c,k}$, denoted as $q_{c,k,0}$ and $q_{c,k,1}$. 
With the three pairs ($\hat{c}_{i,j,k,0}$, $q_{c,k,0}$),  ($\hat{c}_{j,i,k,1}$, $q_{c,k,1}$), and ($\hat{c}_{i,j,k,0}$ $tk_{i,c,k}$), we compute
the bilinear mapping $e$ over each pair and multiply all the results one by one. If the computed result is $1$, the  current  phenotype matches the query, otherwise, it does not match. 
The correctness of the algorithm is shown in Eq. \ref{eq:paper3_correct_tk}.

\alglanguage{pseudocode}
 \begin{algorithm}
  \caption{SinglePhenotypeSearch}
 \label{alg:paper3_search_pairing}
  \small
\begin{algorithmic}[1]
\Require $\hat{c}_{i,j,k}$, $q_{c,k}$, $tk_{i,c,k}$
\Ensure $1$ or $0$
\State  $ (g^{r_{i,j,k}}, c_{i,j,k}, \bar{c}_{i,j,k}) \leftarrow \hat{c}_{i,j,k}$ 
\State  $(\hat{c}_{i,j,k,0},\hat{c}_{i,j,k,1},\hat{c}_{i,j,k,2})
 \leftarrow (g^{r_{i,j,k}}, c_{i,j,k}, \bar{c}_{i,j,k})$ 
\State $ (g^{r_{c,k}}, g^{-r_{c,k}\alpha_{c,k}}PK_c^{r_{c,k}}) \leftarrow q_{c,k}$ 
\State $(q_{c,k,0}, q_{c,k,1}) \leftarrow (g^{r_{c,k}}, g^{-r_{c,k}\alpha_{c,k}}PK_c^{r_{c,k}})$ \If{$e(\hat{c}_{i,j,k,0}, tk_{i,c,k}) e(\hat{c}_{i,j,k,0}, q_{c,k,0})e(\hat{c}_{i,j,k,1}, q_{c,k,1})=1$} 
\State \Return 1
 \Else 
\State \Return 0
 \EndIf
\end{algorithmic}
\end{algorithm}

\begin{equation}
    \begin{split}
        &e(\hat{c}_{i,i,k,0}, tk_{i,c,k}) e(\hat{c}_{i,j,k,0}, q_{c,k,0})e(\hat{c}_{i,j,k,1}, q_{c,k,1}) \\
        &=e(g, g)^{\frac{-r_{i,j,k}r_{c,k}(SK_i-SK_c)}{(SK_i-\alpha_{c,k})}}e(g,g)^{\frac{-r_{i,j,k}r_{c,k}(SK_c-\alpha_{c,k})}{SK_i-\alpha_{i,j,k}}}e(g,g)^{r_{c,k}r_{i,j,k}} \\
        &
        \begin{cases}
        =1 & \alpha_{i,j,k}=\alpha_{c, k} \\
        \neq 1 & otherwise \\
        \end{cases}
    \end{split}
    \label{eq:paper3_correct_tk}
\end{equation}


To support a query that contains  multiple phenotypes,  we extend Algorithm \ref{alg:paper3_search_pairing} to Algorithm \ref{alg:paper3_edbsearch_pairing}. In detail,  
we first initialize two lists $\Upsilon_{\textit{case}}$ 
and $\Upsilon_{\textit{control}}$. Then, for each patient with pseudonym $p_{i,j}$ in hospital $i$'s encrypted phenotype dataset  $\textit{Dict}_i^P$, 
the corresponding encrypted phenotype data $C_{i,j}$ is read. 
Given the encrypted phenotype data $C_{i,j}$, query $q_{c}$, and transform key ($tk_{i,c,1}$ , $\cdots$, $tk_{i,c,n}$),  the per-hospital components of 
the given data
are extracted and passed as the input to the  \textit{SinglePhenotypeSearch} algorithm. 
For example, both the first element of phenotype data query and transform key are extracted, and then input into the \textit{SinglePhenotypeSearch} algorithm. 
If a patient contains all the phenotype data inside a query, the patient's pseudonym $p_{i,j}$ is added to the case group $\Upsilon_{\textit{case}}$. 
If a patient does not match query phenotypes, its pseudonym is added to the control group $\Upsilon_{\textit{control}}$. 
This process is executed until the size of both  case and control groups reaches $N$ or until all the data is processed. 

\alglanguage{pseudocode}
 \begin{algorithm}
  \caption{Search}
 \label{alg:paper3_edbsearch_pairing}
  \small
\begin{algorithmic}[1]
\Require $\textit{Dict}_{i}^P$, $q_{c}$, ($tk_{i,c,1}$,$\cdots$, $tk_{i,c,n}$)
\Ensure {$\Upsilon_{\textit{case}}$ and $\Upsilon_{\textit{control}}$}
\State initialize  two lists $\Upsilon_{\textit{case}}$ 
and $\Upsilon_{\textit{control}}$ 
\State set the number of non-zero elements in $q_c$ as $thr$ 
\State $N \leftarrow q_c$
\ForAll{$ p_{i,j} \in \textit{Dict}_{i}^P$}
\State $\mathbf{C}_{i,j} \leftarrow \textit{Dict}_{i}^P[p_{i,j}]$ 
\ForAll{$\hat{c}_{i,j,k} \in \mathbf{C}_{i,j}$}
\State    \textit{score} $\leftarrow$ \textit{score} + \textit{SinglePhenotypeSearch}($\hat{c}_{i,j,k}$,$q_{c,k}$, $tk_{i,c,k}$) 
\EndFor
\If{score = $thr$ \textit{and} $|\Upsilon_{case}| < N$}
\State $\Upsilon_{\textit{case}}$.\textit{append}($p_{i,j}$)
\ElsIf{score = $0$ \textit{and} $|\Upsilon_{control}| < N$}
\State $\Upsilon_{\textit{control}}$.\textit{append}($p_{i,j}$)
\EndIf
  \If{$|\Upsilon_{\textit{case}}| = N$ and  $|\Upsilon_{\textit{control}}| = N$ }
  \State  \textit{break}
 \EndIf
\EndFor
\State \Return $\Upsilon_{\textit{case}}$ and $\Upsilon_{\textit{control}}$
\end{algorithmic}
\end{algorithm}

\subsection{Computation over Identified Target Patients }

Here, we briefly discuss how our scheme can compute GWAS (or other statistics) over identified patients in centralized and decentralized approaches. 
In a centralized setting, the patient genomic data is also stored at the CSP, and the genomic data of each patient can be encrypted using multi-key fully homomorphic encryption mechanism (e.g., \cite{chen2019efficient}) for storage at the CSP. Then, after identifying the case and control groups (as in Section \ref{sec:paper3_identificationcaseandcontrolgroup}), the CSP can directly execute the GWAS algorithm over the identified patient data using the homomorphic properties of multi-key fully homomorphic encryption mechanism. 
In a decentralized setting, the patient genomic data is not stored at the CSP. In this case, the CSP can return the identified patients to the client. Based on the identified patients, the client can send requests to all the involved hospitals to get access to their data for computing GWAS in a distributed way~\cite{raisaro2018m,froelicher2021truly}.

\subsection{Managing Dynamic Phenotype Data}
\label{sec:paper3_updatedatabase}

 Here, we show how the proposed scheme supports efficient update of patient phenotype data stored at the CSP.  Assume the hospital $i$ wants to update phenotype $k$ of patient $j$. Given the patient pseudonym $p_{i,j}$ and the phenotype ($\textit{P.name}_k, \textit{P.val}_k$), 
the phenotype encryption algorithm (in Section \ref{sec:paper3_dataencryption}) is called to encrypt the phenotype. 
Once the encryption is completed, the vector of the update query is constructed by inserting 
the ciphertext at the position where $\textit{P.name}_k$ is located in $\mathbf{A}$ and setting the remaining values to $0$.
Additionally, the command ``update" is added to the query. The update query is sent to the CSP. Upon receiving the update query, 
the CSP first identifies the entry of the patient $p_{i,j}$, then identifies the location corresponding to the non-zero element inside the vector of the update query, and finally replaces the old cipher with the new cipher from the update query. 
To insert a new phenotype or to delete an existing phenotype from a patient record, the proposed update algorithm can also be used by directly appending or removing a record from stored ciphertext.

\section{Privacy Analysis}
\label{sec:paper3_privacy_analysis}
In this section, we analyze the privacy of phenotype data.
We first provide a high level discussion on how the proposed scheme achieves robustness in the presence of attacks presented in Section~\ref{sec:paper3_security_model}. Then
we  present formal privacy definition and proof.

\subsection{Privacy Against Ciphertext Analysis}
Phenotype data is encrypted and stored at the CSP. This allows the CSP to analyze the stored ciphertext. 
The encrypted phenotype data can be split into two parts (as in Section \ref{sec:paper3_dataencryption_phenotypeenc}). 
The first part of the encrypted phenotype data is constructed in two steps. 
Hospital first computes the hash of the phenotype attribute and value. 
Then, hospital randomly selects a number to mask above hash result and raised the result to the power of a group element. 
The privacy of this part relies on the hardness of discrete logarithm problem, one-wayness of the hash function, and randomness of the selected number.
The second part of the encrypted phenotype data results from directly encrypting phenotype attribute and value. 
Hospital uses its secret key and invokes symmetric encryption algorithm (e.g., AES) to encrypt the concatenation of phenotype attribute and value. 
The privacy of the second part relies on the robustness of the utilized symmetric encryption algorithm (e.g., AES). 
From the above description, we can conclude that both parts of the encrypted phenotype are semantically secure. Thus, the CSP is unable to learn significant information from the ciphertext analysis. 

\subsection{Privacy Against Query Analysis}
The input of the query includes a set of phenotypes. 
Each  phenotype includes a pair of phenotype attribute and  value, which is first hashed and the hash result is lifted as a power of a group element (as in Section \ref{sec:paper3_querygen}). 
Additionally, a random mask is selected to hide this result, which enables the encrypted query to be semantically secure. Due to the semantic security of the query, the CSP is unable to learn the query content from the query analysis.

\subsection{Privacy Against Operation Analysis}
\label{paper3:privacy_operation_analysis}
Since the ciphertext of phenotype genotype data is stored at the CSP, the CSP is responsible for conducting search  over the ciphertext. 
For the search operation, 
the CSP applies the query to search over the ciphertext of phenotype data. If a phenotype is included in the query, the corresponding search result is $1$, otherwise, it is $0$.
Through the execution of the search process, the CSP can learn the number of matching phenotypes of each patient. However, for each matching phenotype, its value can either be 0 or 1, and the CSP  cannot distinguish between the two.
Thus, the CSP cannot learn any information about the query and ciphertext from the search operation. 

\subsection{Robustness Against Unauthorized Access}
Access control is designed through the  \emph{transform key} (as discussed in Section \ref{sec:paper3_clientauthoriatoin}).
A client selects random numbers 
$r_{c,k}$ ($1 \leq k \leq n$) and
sends them to the target hospital. 
The hospital generates the \emph{transform key} ($tk_{i,j,k}=g^{\frac{-r_{c,k}(SK_i-c)}{SK_i-\alpha_{c,k}}}$) by lifting these numbers into the power of a group element. 
Due to the randomness of $r_{c,k}$ and the hardness of the discrete logarithm problem, the CSP is unable to extract any information from the transform key.   
Analyzing the search operation, the output of using incorrect \emph{transform key} is $0$, which does not disclose any information to the CSP, as described in Section \ref{paper3:privacy_operation_analysis}. Therefore, the proposed authorization scheme is robust against the unauthorized access.

\subsection{Robustness Against Colluding Hospitals}
Each hospital independently encrypts its phenotype data and genotype data with its own secret key. 
Even if several hospitals collude with each other, they cannot get any advantage to infer another hospital's data.

\subsection{Robustness against Collusion between a Hospital and CSP}
If one or more hospitals collude with the CSP, the CSP cannot obtain any advantage to infer the remaining hospitals' data since each hospital's data is encrypted independently. 
 However, as we clarified in Section \ref{sec:paper3_security_model}, we do not consider following case. 
 For instance, if the search (at the CSP) includes two hospitals and if one of these hospital collude with the CSP, the CSP can learn which patients of the other (victim) hospital has the considered phenotype as a result of the search operation (but not the genomic data of the identified patients). To provide the common search functionality, this attack is unavoidable, and hence we do not consider it in this work.



\subsection{Privacy Analysis}

 In the following, we provide a formal privacy analysis of the proposed scheme. 
 Following previous works \cite{chen2019efficient}, the allowed leakage includes (i) size pattern and (ii) access pattern. 
The size pattern discloses the size of the ciphertext, while the access pattern reveals the access frequency of matching patient records. The allowed leakage is not considered violation of our privacy goal. The privacy of the proposed scheme is based on the hardness of discrete logarithm problem, the randomness of selected random mask, and the robustness of applied symmetric encryption (e.g., AES).

The privacy of the proposed scheme can be divided into two elements: phenotype data privacy, 
and query privacy.
The privacy of phenotype data can be further divided into two parts. One part is encrypted by using symmetric encryption algorithm (the third element in a ciphertext, detailed in Section \ref{sec:paper3_dataencryption_phenotypeenc}) and the other part is not (the first and second elements in a ciphertext, detailed in Section \ref{sec:paper3_dataencryption_phenotypeenc}).
Due to the robustness of symmetric encryption algorithm (e.g., AES), the part with the symmetric encryption is also semantically secure. The other part is first protected by a hash function and then masked with a random value. Both the hash result and random mask are put into the power of a group element. Due to the random mask and the hardness of discrete logarithm problem, the ciphertext is semantically secure in the presence of chosen plaintext attack. 
The query privacy is achieved similarly, relying on the random mask and discrete logarithm problem.


Formally, we define the privacy experiments as follows. 
Let $\Pi$ be the scheme,  the advantage of the adversary is defined as 
$\text{ADV}_{\mathcal{A}}^\Pi(\lambda)=Pr(b^*=b) -1/2$, where $b$ and $b^*$ are defined below. In the following, we detail the game.

\noindent\textbf{Init}: The adversary $\mathcal{A}$ selects two datasets $DB_0$ and $DB_1$ with same size and sends them to the challenger. 

\noindent\textbf{Setup}: With the input of security parameter $\lambda$, the challenger runs \emph{Setup} 
to initialize the parameters. 
Then, the challenger calls \emph{KeyGen} 
to generate the keys. 

\noindent\textbf{Phase 1}:
$\mathcal{A}$ is allowed to ask the following  request: 
 \begin{asparadesc}
    \item[Phenotype data encryption request:]
    $\mathcal{A}$ is allowed to send a dataset with phenotype data to ask for encryption. The challenger calls \emph{Encrypt} algorithm to encrypt the dataset and sends the result back to $\mathcal{A}$. 
     
  
 \end{asparadesc}

\noindent\textbf{Challenge}:
 The challenger randomly selects $b$ from $\{0, 1\}$, encrypts the dataset $DB_b$, and sends it to the adversary $\mathcal{A}$. 
 
\noindent\textbf{Phase 2}:
$\mathcal{A}$ is allowed to send requests as in
\textbf{Phase 1}. Additionally, $\mathcal{A}$ is allowed to send a \textit{query request}. The challenger only authorizes a query containing the phenotype attributes, where two datasets have the same value. Then, it generates a \emph{transform key} for $\mathcal{A}$, where the mask ($r_{c,k}$) in \emph{transform key} is randomly selected by the challenger (see details in Section \ref{sec:paper3_clientauthoriatoin} ). 

\noindent\textbf{Guess}: $\mathcal{A}$ outputs $b^*$ as the guess for $b$. 

We say the scheme $\Pi$ is privacy-preserving if the  advantage of the adversary  is negligible, i.e, 
$ADV_\mathcal{A}^\Pi \leq negl(\lambda)$, where $negl(\lambda)$ is a negligible function in $\lambda$.

From above defined privacy game, the adversary $\mathcal{A}$ is only allowed to learn the information from \textbf{Phase 1} and \textbf{Phase 2}. 
The difference of \textbf{Phase 2} from \textbf{Phase 1} is that  $\mathcal{A}$ holds the challenged ciphertext and is allowed to ask a query request.  
Thus, we only need to prove that what the adversary $\mathcal{A}$ can learn from ciphertext request and query request is negligible as follows. 
\begin{asparadesc}
    \item[Phenotype data encryption request:]
    $\mathcal{A}$ submits the dataset $DB$ of phenotype data to ask for encryption from hospital $i$. 
    
    If the \emph{PhenotypeEnc} algorithm is semantically secure, $\mathcal{A}$ is unable to distinguish ciphertext from a 
    random string. 
    The ciphertext of each pair of phenotype attribute and value includes three components. The first component $g^{r_{i,j,k}}$ is randomly selected from $\mathbb{Z}_p$, which does not reveal any information. 
    The second component 
    $c_{i,j,k}$ is  $g^{-r_{i,j,k}/(SK_i -\alpha_{i,j,k})}$. Due to the hardness of discrete logarithm problem,
    $\mathcal{A}$ is unable to extract the power of a group element. 
    Similarly,  $\mathcal{A}$
     is unable to distinguish the second component from a random element in $\mathbb{G}$. 
     The third component is encrypted by using a symmetric encryption algorithm (e.g., AES), which is semantically secure. 
    
    Therefore, the ciphertext obtained through \emph{Encrypt} algorithm is semantically secure. 

    
    \item[Query request:]
    First,  the \emph{transform key}  is indistinguishable from a random element of group $G$. 
    Second, for each pair of phenotype attribute and value,
    the query is 
     $(g^{r_{c,k}}$, $g^{-r_{c,k}\alpha_{c,k}}PK_c^{r_{c,k}})$, where 
     $PK_c$ is an element from ${G}$. 
     Based on the hardness of discrete logarithm problem and the randomness of $r_{c,k}$, the query is indistinguishable from two random elements from ${G}$. 
    Third,  
     given the query and \emph{transform key}, the adversary $\mathcal{A}$ is capable to run search operation over the  ciphertext of phenotype data.  Furthermore, $\mathcal{A}$ is also capable to run analysis algorithms on ciphertext of genotype data. 
     However, due to the constraint of issuing client authorization, two datasets should output the same search result. Thus, $\mathcal{A}$ cannot learn any significant information via executing search operation. 
\end{asparadesc}

Based on above analysis, we can conclude 
$ADV^\Pi_{\mathcal{A}}(\lambda)$ is negligible and 
the proposed scheme is privacy-preserving. 

\section{Evaluation}
\label{sec:paper3_evaluation}
In this section, we evaluate the performance of the proposed scheme.  
We run the experiments on a commodity machine with  $i7$ CPU and 16GB RAM.
The proposed phenotype encryption algorithm is implemented by \emph{Charm} \cite{charm13} written in Python 
while
the SNP encryption algorithm is implemented by \emph{HEAAN} \cite{Heaan} written in C++. 
Each experiment is run 10 times; we report the average results. 

\subsection{Data Model}
We use the \emph{rsnps} tool~\cite{rsnps} to obtain all the raw patient files from the publicly available OpenSNP dataset~\cite{OpenSNP}. 
Then, we converted the raw patient files into the VCF format using an open source software named \emph{personal-genome-analysis}~\cite{oga}. Eventually, we ended up with 1052 valid VCF files. 
For the phenotype data, we also used the OpenSNP dataset. In total, we collected 1052 records and extracted 1052 phenotype attributes.

\subsection{Experimental Results}
In this section, we first show 
the efficiency and scalability of the phenotype and genotype data encryption algorithms. 
After that, we present the scalability and efficiency of client authorization and query generation. 

\subsubsection{Phenotype Data Encryption}
We adopt symmetric pairing group \textit{SS512} to construct the bilinear mapping and AES to implement the symmetric encryption. Accordingly, 
the time required to encrypt the phenotype data can be divided into two constituents. The first constituent is due to the pairing group operation. The second constituent is due to using AES to encrypt phenotype attribute and value. The secret key of AES is set to $256$ bits.
Table \ref{tab:phenotypeDataEncryption} shows the performance of the
phenotype data encryption for different number of phenotypes. 
We observed that with the linear increase in the number of patients, the time cost of phenotype encryption for both AES and non-AES (pairing based encryption) constituents increases linearly. 
Similarly, when the number of phenotypes increases, the required encryption time also increases linearly.


 \begin{table}[tb!]
    \caption{Time cost of phenotype data encryption for different number of patients and phenotypes}
    \label{tab:phenotypeDataEncryption}
    \begin{tabular}{|c|c|c|c|}
 \hline 
 \# patients & \# phenotype     &  Non-AES  (s) & AES  (s) \\
 \hline 
 1052   &  1052 & 8137.5 & 5  \\ 
 \hline 
 1052 & 526 & 3954.8 & 2.7 \\
 \hline 
 1052 & 263 & 2068.9 & 1.38 \\
 \hline 
 526 & 1052 & 3948 & 2.7 \\
 \hline 
 263 & 1052 & 2042.1 & 1.38 \\
 \hline
    \end{tabular}
\end{table}

 \begin{table}[tb!]
     \caption{Time cost of authorization request generation for different number of requested phenotypes.}
    \label{tab:authorisationrequestgen}
    \begin{tabular}{|c|c|}
    \hline
     \# requested phenotypes    & time (s) \\
    \hline
  1052      &  4.47 \\
  \hline 
  526 & 2.28 \\
  \hline 
  263 & 1.14 \\ 
  \hline 
   \end{tabular}
\end{table}

\subsubsection{Client Authorization}
The process of client authorization can be divided into two phases. 
In the first phase, the client generates
an authorization request while in the second phase, the hospital generates the \textit{transform key}. Table
\ref{tab:authorisationrequestgen} shows the efficiency of authorization request generation for different number of requested phenotypes. 
We observe that
the required time of authorization request generation increases linearly upon increasing the number of requested phenotype attributes. 
Table \ref{tab:clientAuthorization} shows the performance of the \textit{transform key} generation for different number of phenotype attributes. Here, we observe that the time required for \textit{transform key} generation increases linearly as a function of the number of phenotype attributes.


\begin{table}[t!]
      \caption{Time cost of \textit{transform key} generation for different number of authorized phenotypes}
    \label{tab:clientAuthorization}
    \begin{tabular}{|c|c|}
    \hline 
  \# authorized phenotypes & time (s) \\
  \hline 
  1052      &  5.87 \\
  \hline 
  526 & 3.2 \\
  \hline 
  263 & 1.62 \\ 
  \hline 
    \end{tabular}
\end{table}
\begin{table}[t!]
      \centering
          \caption{Time cost of query generation for different number of input phenotypes}
    \label{tab:queryGeneration}
    \begin{tabular}{|c|c|}
    \hline 
  \# input   phenotypes  & time (s) \\
  \hline 
  1052      &  4.8 \\
  \hline 
  526 & 2.39 \\
  \hline 
  263 & 1.19 \\ 
  \hline 
    \end{tabular}
\end{table}


\subsubsection{Query Generation}
The performance of the query generation algorithm is affected by the number of input phenotypes. 
The experimental results are shown in Table \ref{tab:queryGeneration}. 
From the table, we observe that with the linear increase in the number of input phenotypes, the time required for the query generation also increases linearly.


\subsubsection{Phenotype Data Search}
\label{sec:paper3_evaluation_phenotype_search}
The search process involves a number of patient records to be processed to construct the case and control groups. 
To access each hospital's data, the query needs to be transformed using the \textit{transform key}. 
In addition, the desired size of case and control groups can be a factor to stop the search process earlier once the required number of individuals are identified in case and control groups. 
Table \ref{tab:phenotypeSearch} shows the effect of the number of hospitals, number of input phenotypes, and the size of case and control groups on the efficiency. 
We observed that if the number of patients is fixed, the number of hospitals almost does not affect the efficiency, while the performance is sensitive to the size of case and control groups and the number of input phenotypes.  
When the size of case/control groups are set to $10$ and $50$, the search time is reduced to $0.29$s and $1.78$s, respectively. The reason is that once the required number of patients are identified for the case and control groups, the search process is terminated.
We also evaluate the performance for reduced number of input phenotypes. The observed search times are $32.2$s, 
$166.7$s, and $327.1$s for $10$, $50$, and $100$ phenotypes, respectively.
From these results,  we can say that the search time increases linearly as the number of input phenotypes grows.

\begin{table*}[htb]
    \centering
     \caption{Time cost of phenotype data search for the proposed algorithm with a total of 1052 patients and each patient having 1052 phenotypes}
    \label{tab:phenotypeSearch}
    \begin{tabular}{|c|c|c|c|c|}
    \hline 
\# hospitals  &  \# queried  phenotypes  & \# case/control groups & time (s)\\
    \hline 
1  & 10 & 100 & 32.2 \\
\hline 
10  & 10 & 100 & 33.1 \\
\hline 
100  & 10 & 100 & 34.7 \\
\hline 
1  & 10 & 10 & 0.29 \\
\hline 
1  & 10 & 50 & 1.78 \\
\hline 
1  & 50 & 100 & 166.7 \\
\hline 
1  & 100 & 100 &327.1 \\
\hline 
\end{tabular}
\end{table*}

To show the efficiency of the proposed algorithm for phenotype data search, we also designed a following fully homomorphic encryption (FHE)-based version of it for comparison. 
In detail, the alternative FHE-based approach includes four steps.
First, all the phenotype data is encrypted by using CKKS \cite{cheon2017homomorphic}. 
Second, the client sends a query to the CSP in order to determine the case/control groups. 
Third, the CSP sends the computed result to different hospitals. 
Fourth, hospitals decrypt the result in parallel and send the result (identified case/control groups) to the client. 
Without considering the time cost of communication, Table \ref{tab:phenotypeSearch_FHE} shows the required time to complete the search operation using this FHE-based scheme for different number of patients and phenotypes. 
Comparing Table \ref{tab:phenotypeSearch} with Table \ref{tab:phenotypeSearch_FHE}, we observe that the proposed algorithm is more than 20 times faster than the FHE-based algorithm when 
the number of queried phenotypes is at most 10.
Notably, when the size of case/control groups is smaller than 50, 
the proposed algorithm is more than 300 times faster.
Furthermore, in Table \ref{tab:phenotypeSearch_FHE_patients} we show the comparison between the FHE-based solution and proposed algorithm for different number of patients. 
From the table, we observe that (i) the proposed scheme consistently exhibits similar performance advantage over the FHE-based scheme and (ii) the time cost of both algorithms increases linearly with the number of patients.

\begin{table*}[t!]
    \centering
     \caption{Time cost of phenotype data search for the CKKS algorithm with a total of 1052 patients and each patient having 1052 phenotypes}
    \label{tab:phenotypeSearch_FHE}
    \begin{tabular}{|c|c|c|c|c|}
    \hline 
\# hospitals  &  \# queried  phenotypes  & \# case/control group & each hospital decryption time (s) & total \\
    \hline 
1  & 10 & 100 & 6.2 & 684.6\\
\hline 
10  & 10 & 100 & 5.9 & 683.9\\
\hline 
100  & 10 & 100 & 0.6 & 678.6\\
\hline 
1  & 10 & 10 &  1.5 & 678.5\\
\hline 
1  & 10 & 50 &  3.1 & 681.1\\
\hline 
1  & 50 & 100 &  6.3 & 684.7\\
\hline 
1  & 100 & 100 & 6.5 & 684.9\\
\hline 
    \end{tabular}
\end{table*}

\begin{table}[t!]
    \centering
     \caption{Time cost of phenotype data search for the CKKS and proposed algorithm for different number of patients (each patient having 1052 phenotypes). The size of case/control groups is not limited as input parameter.}
    \label{tab:phenotypeSearch_FHE_patients}
    \begin{tabular}{|c|c|c|c|}
    \hline 
\# hospital  &  \# queried   phenotypes & \# patients&  total time (s)\\
    \hline 
\multicolumn{4}{|c|}{CKKS algorithm} \\
\hline
1  & 10 & 1052  & 747.5\\
\hline 
1  & 10 & 526  & 355.9 \\
\hline 
\multicolumn{4}{|c|}{Proposed algorithm}\\
\hline
1  & 10 & 1052 & 35.9\\
\hline 
1  & 10 & 526 & 17.4\\
\hline 
    \end{tabular}
   
\end{table}

\section{Discussion}
\label{sec:extension_proposed_scheme}
The proposed pairing-based encryption scheme for phenotype data (in Section~\ref{sec:paper3_identificationcaseandcontrolgroup}) is not limited to efficient identification of case and control groups. The scheme can be extended to support additional functions, such as similar patient search, target record retrieval, and statistical analysis of phenotype data. In the following, we discuss some of these potential extensions.

\noindent\textbf{Similar patient search.} In the proposed scheme, a client is allowed to input several phenotypes of interest into a query and send it to the CSP. 
Upon receiving the query, the CSP searches the stored datasets of several hospitals. 
Based on the number of matching phenotypes, the CSP can order the search results and return similar patient records. Similar functionality can also be provided in the genome level  by encrypting genome data with the proposed pairing-based encryption.

\noindent\textbf{Phenotype data retrieval.} In the proposed scheme, we show how to use the \emph{Search} algorithm (in Section~\ref{sec:paper3_identificationcaseandcontrolgroup}) to find  target pseudonyms based on the query. 
One can extend this function to support a client to retrieve phenotype data of interest. 
For instance, a client (e.g., a physician) may be interested to know the phenotypes of patients having lung cancer. Then, the client can generate a query and send it to the CSP to retrieve the phenotype data from patients that are diagnosed with lung cancer. 
Once the client receives the phenotype data from the CSP, the \emph{SinglePhenotypeDecrypt} algorithm \ref{alg:paper3_decrypt_pairing} is called to decrypt the phenotype data. 

\alglanguage{pseudocode}
 \begin{algorithm}
  \caption{SinglePhenotypeDecrypt}
 \label{alg:paper3_decrypt_pairing}
  \small
\begin{algorithmic}[1]
\Require $\hat{c}_{i,j,k}$, $K_i$
\Ensure $\textit{P.name}_k$, $\textit{P.val}_{j,k}$
\State $ (g^{r_{i,j,k}}, c_{i,j,k}, \bar{c}_{i,j,k}) \leftarrow \hat{c}_{i,j,k}$
\State
 $(\textit{P.name}_k, \textit{P.val}_{j,k}) \leftarrow DE(K_i, \bar{c}_{i,j,k})$ 
 
\State \Return  $\textit{P.name}_k, \textit{P.val}_{j,k}$
 \end{algorithmic}
\end{algorithm}
\section{Conclusion}\label{sec:conclusion}
In this paper, we have designed a privacy-preserving framework for 
identification of a target group of patients
across multi-tenant data.  
To achieve this, we have proposed a novel phenotype encryption algorithm. 
To support search and computation over multi-tenant data by a cloud service provider (CSP), we have introduced a \emph{transform key} to enable the CSP to transform a single query and execute it over different hospitals' datasets without privacy violation. 
To manage the authorization of clients, we have proposed a per-query based authorization mechanism supporting selective phenotype data authorization.
Via  simulations  on  real genomic data, we have shown the practicality and efficiency of the proposed scheme. 
We believe that the proposed scheme will  further  facilitate  the  use  of  genomic  data  in  clinical settings  and  pave  the  way  for  personalized  medicine.  In future work, we will focus on improving the search efficiency of genomic data and batch queries.

\bibliographystyle{splncs04}
\bibliography{main}

\appendix


\end{document}